\documentclass[10pt,twocolumn,letterpaper]{article}

\usepackage{wacv}              

\usepackage[pagebackref,breaklinks,colorlinks]{hyperref}

\usepackage{orcidlink}

\usepackage[utf8]{inputenc} 
\usepackage[T1]{fontenc}    
\usepackage{hyperref}       
\usepackage{url}            
\usepackage{booktabs}       
\usepackage{amsfonts}       
\usepackage{amsmath}        
\usepackage{amsthm}         
\usepackage{nicefrac}       
\usepackage{microtype}      
\usepackage{xcolor}         
\usepackage{graphicx}

\usepackage{booktabs}   
\usepackage{amssymb}    
\usepackage{amssymb}
\usepackage{mathtools}
\definecolor{cvprblue}{rgb}{0.21,0.49,0.74}

\usepackage{natbib}
\usepackage{tikz}
\usepackage{quantikz}
\usetikzlibrary{positioning, backgrounds,  arrows.meta, shapes.geometric,  fit,  shapes, arrows.meta }

\tikzset{
  smallarrow/.style={
    -{Stealth[length=1mm,width=0.9mm]}
  }
}

\usepackage{algorithm}
\usepackage{algpseudocode}

\usepackage{colortbl} 
\definecolor{wacvblue}{rgb}{0.21,0.49,0.74}

\def\wacvPaperID{1956} 
\def\confName{WACV}
\def\confYear{2027}

\title{QiT: Quantum-Inspired Transformer for Visual Recognition Task}

\author{Badri N. Patro\\
Microsoft\\
\and
 Vijay S. Agneeswaran \\
Microsoft\\
}

\begin{document}
\maketitle

\begin{abstract}

Quantum machine learning offers a compelling representational perspective: angle-encoded states inhabit Hilbert spaces in which periodic similarities and interactions can be expressed naturally. Realizing this perspective for visual recognition remains difficult, however, because present quantum neural networks are constrained by limited qubit counts, costly circuit simulation and measurement, noise, and unstable optimization on noisy intermediate-scale quantum devices. We investigate whether useful structural ideas from quantum models can instead be realized as scalable classical Transformer operations. We introduce QiT, a Quantum-inspired Transformer for vision tasks with three components: (i) angle-inspired encoding that maps image tokens to learned trigonometric Hilbert-space features analogous to quantum rotation-based state encoding; (ii) self-attention over these periodic features, inducing a classical cosine kernel approximated to quantum fidelity kernels; and (iii) gated multiplicative emulation, a trainable classical surrogate for interaction terms found in variational circuits. All components are differentiable tensor operations, so QiT claims neither quantum computation nor quantum speedup and retains the $\mathcal{O}(N^2D)$ attention complexity of a standard Vision Transformer. Across image-classification benchmarks, QiT is competitive with a matched classical Transformer while avoiding the severe runtime cost observed for a small simulated quantum Transformer. QiT-B reaches 78.3\% ImageNet-1K top-1 accuracy with 45.7M parameters and 11.5 GFLOPs. These results position QiT as a scalable baseline for isolating and evaluating quantum-motivated inductive biases in visual recognition.

\end{abstract}

\section{Introduction}
\label{sec:intro}

Self-attention gives vision models a direct mechanism for relating distant image patches~\cite{NIPS2017_Vaswani,dosovitskiy2020image}. Its query and key projections are linear, however, so nonlinear pairwise structure must emerge through depth and the surrounding feed-forward blocks. Quantum machine learning suggests a complementary design vocabulary: periodic angle encodings, feature-space inner products, and multiplicative interactions produced by parameterized quantum circuits (PQCs)~\cite{Biamonte_2017, QML_Hilbert_Space_19}. Whether these ingredients are useful in a scalable classical vision model remains insufficiently separated from the question of whether a model runs on quantum hardware.

Current quantum neural networks face three obstacles to vision-scale evaluation. First, recurrent designs such as QRNNs~\cite{bausch2020recurrent} and QLSTMs~\cite{chen2022quantum} process tokens sequentially, limiting parallelism and making direct long-range interaction less efficient than self-attention. Second, fully quantum and hybrid Transformers remain difficult to scale on noisy intermediate-scale quantum (NISQ) hardware~\cite{Preskill_2018}: available qubit counts, feasible circuit depths, measurement noise, and repeated sampling constrain both model size and evaluation throughput~\cite{cherrat2024qvt, Zhao_2024_QKSAN,li2023quantumselfattentionneuralnetworks}. Third, some parameterized circuit families develop barren plateaus, in which gradient concentration impedes optimization as the system grows~\cite{mcclean2018barren}. Existing hybrid approaches face a dilemma: light hybrids delegate only a few layers to quantum circuits and gain little expressivity, while heavy hybrids attempt full quantum attention but cannot train at realistic sequence lengths. Consequently, quantum-attention studies are generally evaluated at substantially smaller scales than modern vision Transformers, and controlled large-scale comparisons remain scarce.

\begin{figure*}
\centering
\includegraphics[width=0.949\textwidth]{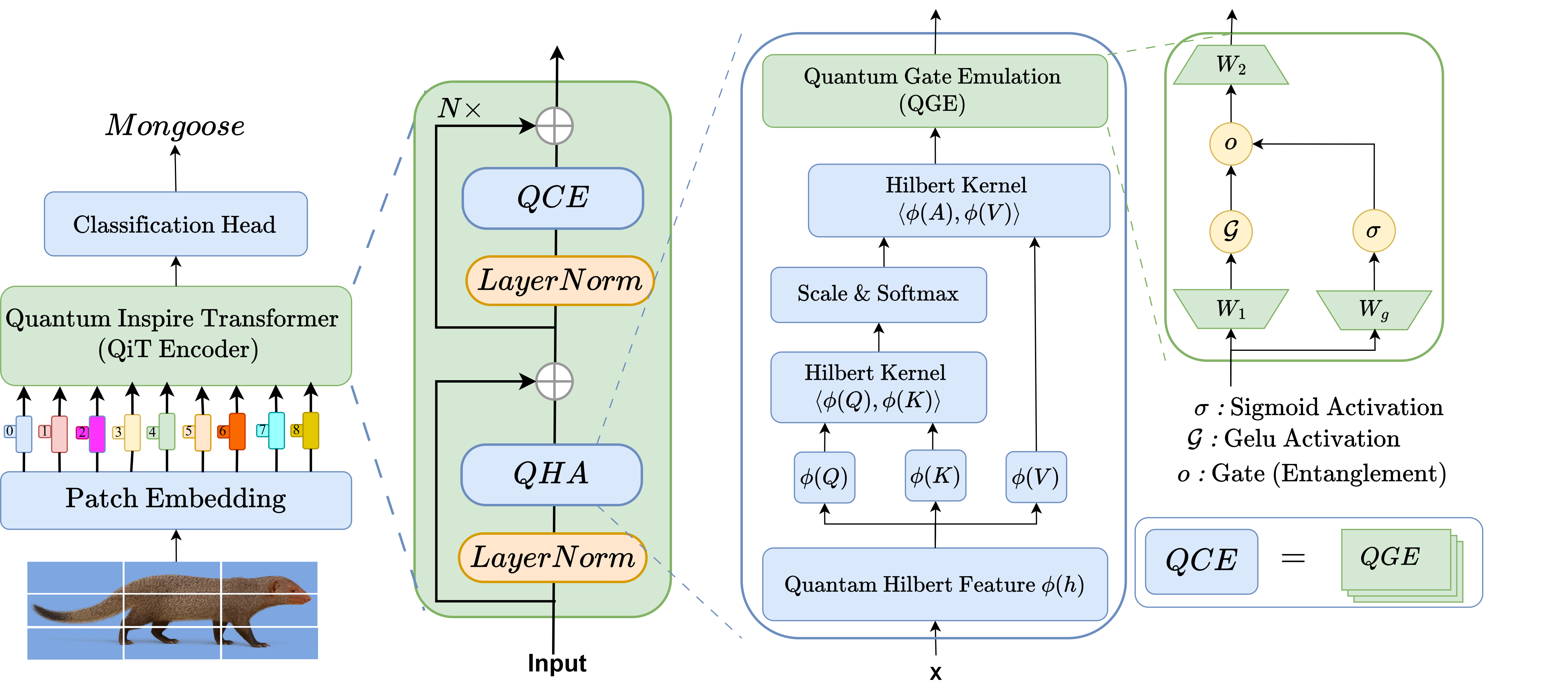}
\vspace{-0.15in}
\caption{\textbf{QiT encoder block: Tokens are mapped to a quantum-inspired Hilbert space via angle encoding and Fourier features. Kernel attention approximates quantum fidelity, while local quantum mixers and a circuit surrogate layer emulate bounded-depth PQCs within a Transformer architecture. } }
\label{fig:main}
\vspace{-0.25in}
\end{figure*}

This work asks whether structural motifs associated with quantum models in Hilbert spaces can provide useful inductive biases in a Transformer that retains the scalability of classical deep learning. More precisely, can periodic angle features and multiplicative interactions be integrated into parallel visual attention without circuits, measurements, or assumptions of quantum computational advantage? We need an architecture that integrates quantum principles without quantum hardware. Figure~\ref{fig:main} provides an overview of the resulting QiT encoder.

We propose QiT (Quantum-inspired Transformer), a classical vision architecture that translates three motifs from parameterized quantum models into differentiable tensor operations. First, \emph{angle-inspired encoding} maps each token $h$ to bounded angles $\theta(h)=\pi\tanh(hW_\theta+b)$ and trigonometric periodic features $\phi(h)=[\sin(\theta(h)),\cos(\theta(h))]$, analogous in form to quantum rotation-based data encoding. Second, \emph{periodic-feature self-attention} forms queries, keys, and values from learned projections of $\phi(h)$ in Hilbert-inspired feature space, allowing tokens to interact after a nonlinear periodic lift. Before projection, the inner product of these features is an explicit cosine kernel; after learned query and key projections, the attention score is a trainable bilinear similarity rather than a quantum fidelity kernel. Third, a \emph{gated multiplicative mixer}, $(\operatorname{GELU}(ZW_1)\odot\sigma(ZW_g))W_2$, introduces input-dependent channel interactions motivated by multiplicative terms in a variational quantum circuit (VQC). It does not instantiate or simulate physical entanglement.

For image tokens $X\in\mathbb{R}^{B\times N\times D}$ and labels $y$, QiT solves standard supervised classification by minimizing cross-entropy for $f_\Theta(X)$. The design reverses the usual hybrid construction: rather than executing classical attention through quantum circuits, it isolates selected quantum-motivated functional forms and evaluates them with conventional GPU kernels. QiT consequently preserves parallel token processing, standard backpropagation, and the $\mathcal{O}(N^2D)$ attention complexity of a Vision Transformer while avoiding circuit simulation, repeated measurement, and quantum-hardware constraints. The resulting model enables these inductive biases to be evaluated on vision benchmarks at a scale that is currently impractical for circuit-based attention.

\paragraph{Contributions.}
\begin{enumerate}
    \item We introduce QiT, a vision Transformer that integrates learned angle-inspired features, self-attention over their quantum-inspired Hilbert-space embeddings, and a gated multiplicative channel mixer. QiT requires no quantum hardware and retains the parallel execution and $\mathcal{O}(N^2D)$ attention complexity of a standard Vision Transformer.
    
    \item We characterize the trigonometric feature map as an explicit positive-semidefinite cosine kernel, connecting QiT to periodic feature methods and rotation-based quantum data encoding. We state the boundary of this connection explicitly: learned query--key scores are bilinear similarities, not general quantum fidelity kernels, and gated mixing does not realize physical entanglement.
    
    \item We evaluate QiT across image-classification benchmarks and compare it with both classical vision models and a four-qubit PennyLane Transformer. In the controlled one-epoch simulator study, QiT reduces runtime by nearly three orders of magnitude; at scale, QiT-B reaches 78.3\% ImageNet-1K top-1 accuracy with 45.7M parameters and 11.5 GFLOPs.

\end{enumerate}

\section{Related Work}\label{sec:related_work}
\paragraph{Transformers and Long-Range Sequence Modeling.}
Transformers enable parallel, content-dependent token interaction across language and vision~\cite{NIPS2017_Vaswani,dosovitskiy2020image,guo2022cmt}. Vision alternatives include spectral token mixers~\cite{rao2023gfnet,patro2025spectformer,patro2023scattering} and state-space models~\cite{gu2023mamba,patro2024simba,liu2024vmamba,yang2024plainmamba, Patro_2026_CVPR_hamsa, Patro_2026_CVPR_nakul}. QiT retains softmax self-attention and changes the feature map supplied to its projections and the subsequent channel mixer.

\paragraph{Quantum Machine Learning.}
Quantum machine learning (QML) encodes classical data into quantum states, applies parameterized quantum circuits (PQCs), and measures observables to produce classical outputs~\cite{Biamonte_2017, QML_Hilbert_Space_19, Havlek2018SupervisedLW}. A register of $Q$ qubits represents states in a $2^Q$-dimensional Hilbert space, but this dimensionality alone does not imply an efficiently accessible learning advantage. Quantum feature maps can induce kernels through state overlaps and generate input-dependent interactions whose utility depends on the encoding, circuit, measurement, and data~\cite{Havlek2018SupervisedLW,fellner2025quantum}. Variational quantum algorithms combine parameterized circuits with classical optimization, often favoring shallow circuits for near-term devices~\cite{mitarai2018quantum,cerezo2021variational,Cerezo_2021_VQA}. Their practical scale remains constrained by qubit availability, gate errors, decoherence, and finite-shot estimation on noisy intermediate-scale quantum hardware~\cite{Preskill_2018}. Moreover, barren plateaus can cause gradients to concentrate near zero for particular circuit families, objectives, and initialization regimes~\cite{mcclean2018barren}.

\paragraph{Quantum Deep Learning for Sequential Data.}
Quantum deep learning develops trainable quantum analogues of neural layers and studies their expressivity and optimization~\cite{mitarai2018quantum,beer2020training,abbas2021power,kwak2021quantum,qian2022dilemma}. For sequential data, QRNNs and related recurrent models update quantum registers step by step~\cite{bausch2020recurrent,li2023qrnn,takaki2021qrnn,li2024qgrnn}; QLSTMs similarly incorporate PQCs into recurrent gating structures~\cite{chen2022quantum,shen2017quantum}. Linear-layer variants and other hybrid modifications seek more stable or efficient training~\cite{cao2023linear}. Data re-uploading and unitary or variational constructions provide additional routes to expressive circuit functions~\cite{rivera2022time,schuld2021effect,perez2020data,shin2023exponential}. These models demonstrate that PQCs can process ordered data, but recurrent execution remains sequential and does not provide Transformer-style all-pairs token interaction. QiT instead studies parallel interactions among image patches and executes all operations classically.

\paragraph{Quantum Attention and Quantum Transformers.}
Quantum attention methods differ in which Transformer operations are assigned to circuits. Light hybrid models such as QSANN~\cite{li2023quantumselfattentionneuralnetworks} and QSAM~\cite{qsam2023} use circuits for selected projections while retaining classical attention operations. More circuit-intensive approaches, including QViT~\cite{cherrat2024qvt} and QKSAN~\cite{Zhao_2024_QKSAN}, incorporate quantum subroutines or kernels into visual attention, increasing circuit-evaluation and measurement demands. QSAN~\cite{shi2024qsan} and F-QSANN~\cite{zheng2023fqsann} pursue fully quantum formulations, whereas HQViT~\cite{hqvit2025} and quantum self-attention for molecular generation~\cite{smaldone2025hybrid} investigate other hybrid allocations. Their encodings, circuit access assumptions, sampling budgets, and output readouts differ, so asymptotic quantities are not uniformly comparable and do not by themselves establish end-to-end speedups.

QiT occupies a distinct point in this landscape: it assigns no operation to quantum hardware. Instead, it translates angle-inspired periodic encoding and circuit-motivated multiplicative interactions into classical tensor operations, then applies standard softmax attention. This design enables conventional GPU training at vision scale, but its efficiency is classical and should not be interpreted as quantum computational advantage.

\begin{table*}
\centering
\caption{Reported or asymptotic resource characteristics of quantum-attention designs. Values follow different circuit, oracle, sampling, and readout assumptions and are therefore descriptive rather than directly comparable. $N$ denotes the number of tokens, $D$ the embedding dimension, and $Q$ the number of qubits. PQG denotes parameterized quantum gates.}
\label{tab:quantum_complexity}
\vspace{-0.9em}
\resizebox{\textwidth}{!}{
\begin{tabular}{l|c|c|c|c|c|c|c}
\toprule
\textbf{Model} 
& \textbf{\#Qubits} 
& \textbf{\#PQG} 
& \textbf{\#Distinct Circuits} 
& \textbf{\#Total Measurements} 
& \textbf{Quantum Outputs} 
& \textbf{Reported Classical Cost} 
& \textbf{Execution Setting} \\
\midrule

CViT (Classical) 
& 0 
& 0 
& 0 
& 0 
& -- 
& $O(N^2 D)$ 
& Classical GPU \\

\midrule

QSANN~\citep{li2023quantumselfattentionneuralnetworks}
& $O(D)$ 
& $O(D)$ 
& $O(N)$ 
& $O(ND)$ 
& $\{q_i\}_{i=1}^{N},\{k_i\}_{i=1}^{N},\{v_i\}_{i=1}^{N}$ 
& $O(ND^2)$ 
& Hybrid/simulated \\

QSAM~\citep{qsam2023}
& $O(D)$ 
& $O(D)$ 
& $O(N)$ 
& $O(ND)$ 
& $\{q_i\}_{i=1}^{N},\{k_i\}_{i=1}^{N},\{v_i\}_{i=1}^{N}$ 
& $O(ND^2)$ 
& Hybrid/simulated \\

HQViT~\citep{hqvit2025}
& $O(\log ND)$ 
& $O(\log D)$ 
& $O(1)$ 
& $O(N^2 \log D)^{*}$ 
& $\{A_{ij}\}_{i,j=1}^{N},\{v_i\}_{i=1}^{N}$ 
& $O(N^2 D)$ 
& Hybrid proposal \\

QViT~\citep{cherrat2024qvt}
& $O(N + D)$ 
& $O(D \log D)$ 
& $O(N)$ 
& $O(ND)^{*}$ 
& $\{y_i\}_{i=1}^{N}$ 
& $O(N^2 D)$ 
& Hybrid/simulated \\

QKSAN~\citep{Zhao_2024_QKSAN}
& $O(D)$ 
& $O(D)$ 
& $O(N^2)$ 
& $O(N^2 D)^{*}$ 
& $\{v_i\}_{i=1}^{N}$ 
& $O(N^2 D)$ 
& Hybrid/simulated \\

QSAN~\citep{shi2024qsan}
& $O(N\log D + N^2)$ 
& $O(N\log D)$ 
& $O(1)$ 
& $O(\log D)$ 
& $\mathbf{y}$ 
& $O(N^2 D)$ 
& Fully quantum proposal \\

F-QSANN~\citep{zheng2023fqsann}
& $O(N\log D)$ 
& $O(N^2 \log D)$ 
& $O(1)$ 
& $O(1)$ 
& Class probability 
& $O(N^2 D)$ 
& Fully quantum proposal \\

\midrule

QiT (Ours) 
& 0
& 0
& 0
& 0
& Classical features
& None
& \textbf{Classical GPU} \\
\bottomrule
\end{tabular}}
\vspace{-1mm}
\end{table*}

\paragraph{Positioning of QiT.}
QiT is neither a quantum Transformer nor a simulator of a general PQC. It isolates three design components---learned periodic token features, attention over their projections, and gated multiplicative channel mixing---within a classical Transformer. The unprojected feature inner product is an explicit cosine kernel, whereas the learned query--key score is generally neither a positive-semidefinite kernel nor quantum-state fidelity. Likewise, multiplicative gating supplies trainable feature interactions but does not create physical entanglement. This positioning makes the quantum connection precise enough to test while permitting direct GPU evaluation. Unlike fully quantum architectures, QiT does not require quantum hardware, avoids barren plateaus and measurement noise, and scales to long sequences and large datasets. Unlike prior hybrid models that use PQCs as black-box nonlinear layers or rely on quantum kernel estimation. The small PennyLane experiment compares practical execution costs in a controlled setting; the large-scale vision results evaluate QiT as a classical architecture rather than evidence of quantum advantage.



\section{Quantum-Inspired Transformer Architecture (QiT)}
\label{sec:method}

QiT is a classical Transformer that uses periodic token features and gated multiplicative channel mixing. These choices are motivated by angle encoding and interaction terms in parameterized quantum circuits (PQCs), but QiT neither executes nor generally simulates a PQC. We first define the implemented operations and then state the limited sense in which they relate to quantum models.

\subsection{Background: Classical Multi-Head Self-Attention}

Given token embeddings $H\in\mathbb{R}^{B\times L\times d}$ (batch size $B$, sequence length $L$, embedding dimension $d$), classical multi-head self-attention computes:
\begin{align}
  Q &= HW_Q, \quad K = HW_K, \quad V = HW_V,\\
  A &= \mathrm{softmax}\!\left(\frac{QK^\top}{\sqrt{d_h}}\right), \quad \mathrm{Attn}(H)= AV,
\end{align}
where $W_Q,W_K,W_V\in\mathbb{R}^{d\times d_h}$ project to queries, keys, and values, and $d_h$ is the per-head dimension. Multi-head attention parallelizes this across $n_h$ heads with independent parameters, concatenating outputs for expressivity. This mechanism enables \emph{direct pairwise token interactions} $i\leftrightarrow j$ without sequential processing, \emph{parallel computation} via batched matrix operations, and \emph{flexible multivariate integration} through embeddings—advantages over MLPs (no temporal bias without engineered features), CNNs/TCNs (indirect long-range modeling requiring depth), and RNNs/LSTMs (sequential bottlenecks, gradient degradation). However, attention incurs $\mathcal{O}(L^2)$ cost and standard linear projections lack the high-dimensional feature space expressivity promised by quantum approaches.

\subsection{Token Embedding (Classical Preprocessing)}

Given an input sequence $X \in \mathbb{R}^{B \times L \times D}$ with batch size $B$, sequence length $L$, and input dimension $D$, we construct token embeddings $H \in \mathbb{R}^{B \times L \times d}$ using standard embedding layers that combine value embeddings, positional encodings, and optional temporal covariates. This preprocessing step follows conventional Transformer practice and costs $\mathcal{O}(BLD)$ operations. The quantum-inspired components begin after embeddings are formed.

\subsection{Quantum Angle Encoding}

In quantum computing, angle encoding prepares quantum states by rotating qubits according to input features. We emulate this classically by mapping each token embedding $h \in \mathbb{R}^d$ to rotation angles:
\begin{equation}
\boldsymbol{\theta}(h) = \pi \tanh(Wh + b), \qquad \boldsymbol{\theta}(h) \in \mathbb{R}^{Q},
\end{equation}
where $Q$ is the number of learned angles (virtual qubits) and $W \in \mathbb{R}^{Q \times d}$ and $b \in \mathbb{R}^{Q}$ are trainable parameters. We use $Q$ to retain the connection to a $Q$-qubit angle encoding, although no qubits are instantiated. The $\tanh$ nonlinearity bounds the angles to $[-\pi,\pi]$ matching the periodicity of quantum rotation gates. We then construct a trigonometric feature map in Hilbert space:
\begin{equation}
\phi(\mathbf{h}) = \frac{1}{\sqrt{Q}}[\sin(\boldsymbol{\theta}(\mathbf{h})), \cos(\boldsymbol{\theta}(\mathbf{h}))] \in \mathbb{R}^{2Q}.
\end{equation}
This embedding corresponds to a Fourier feature representation and induces an implicit positive-semidefinite kernel:
\begin{equation}
k(\mathbf{h}, \mathbf{h}') = \langle \phi(\mathbf{h}), \phi(\mathbf{h}') \rangle
= \frac{1}{Q}\sum_{q=1}^{Q}\cos(\theta_q(\mathbf{h})-\theta_q(\mathbf{h}')).
\end{equation}
Positive semidefiniteness follows directly because $k$ is an inner product. The map is computed with classical trigonometric functions at cost $\mathcal{O}(BLdQ)$. For restricted product-state angle encodings, related trigonometric terms occur in state overlaps; this observation does not make $k$ equal to the fidelity kernel of a general entangling circuit.

\paragraph{Connection to quantum feature maps.}
Angle-encoded quantum models also produce periodic functions of their inputs. QiT retains this periodic structure in a $2Q$-dimensional real feature space, rather than representing a $2^Q$-dimensional state vector. Its connection to quantum feature maps is therefore architectural and spectral, not an equivalence between the represented state spaces.

\subsection{Hilbert-Space Self-Attention}

Standard Transformer attention computes query-key similarities in the original embedding space. QiT instead performs attention in the quantum-inspired feature space defined by angle encoding. For each token $h_i \in \mathbb{R}^d$, we first apply the trigonometric mapping:
\begin{align}
    \theta_i &= \pi\,\tanh(Wh_i + b) \in \mathbb{R}^{Q},\\
  \phi(h_i) &= [\sin\theta_i,\ \cos\theta_i] \in \mathbb{R}^{2Q}.
\end{align}
We then form queries, keys, and values by projecting these features:
\begin{align}
  q_i &= \phi(h_i)W_Q,\quad k_i=\phi(h_i)W_K,\quad v_i=\phi(h_i)W_V,
\end{align}
where $W_Q, W_K, W_V \in \mathbb{R}^{2Q \times d_h}$ are learnable projection matrices and $d_h$ is the per-head dimension. Attention weights follow the standard scaled dot-product formula:
\begin{equation}
A_{ij}
=
\frac{\exp\left(q_i^\top k_j / \sqrt{d_h}\right)}
{\sum_{j'} \exp\left(q_i^\top k_{j'} / \sqrt{d_h}\right)},
\end{equation}
and the attention output is:
\begin{equation}
\mathrm{Attn}(Q,K,V) = A V.
\end{equation}

\begin{figure}[t]
\centering
\scriptsize
\begin{tikzpicture}[
    >=Stealth,
    thick,
    node distance=1.1cm,
    every node/.style={inner sep=2pt}
]
\node (x) [inner sep=1pt, outer sep=0pt] {$x \in \mathbb{R}^d$};

\node (quantum) [right=of x] {$|\psi(x)\rangle = U_E(x)|0\rangle$};
\node (fourier) [below=of quantum] {$\phi(x) = [\sin \theta(x), \cos \theta(x)]$};

\node (kernelQ) [right=of quantum] {$K_Q(x,x') = |\langle \psi(x)|\psi(x')\rangle|^2$};
\node (kernelF) [right=of fourier] {$K_m(x,x') = \langle \phi(x),\phi(x')\rangle$};

\draw[-{Stealth[length=1.6mm,width=1.1mm]}]
  (x) -- (quantum)
  node[midway, above] {\scriptsize Angle}
   node[midway, below] {\scriptsize Encoding};

\draw[->, dashed] (quantum) -- node[left] {\scriptsize Classical Surrogate} (fourier);

\draw[-{Stealth[length=2mm,width=1.2mm]}]
  (quantum) -- (kernelQ)
  node[midway, above] {\scriptsize Fidelity}
  node[midway, below] {\scriptsize Kernel};

\draw[-{Stealth[length=2mm,width=1.2mm]}]
(fourier) -- (kernelF)
  node[midway, above] {\scriptsize Fourier}
  node[midway, below] {\scriptsize Kernel};

\draw[->, double] (kernelQ) -- node[right] {\scriptsize $m\to\infty$} (kernelF);

\end{tikzpicture}
\caption{Two distinct routes from angle-encoded inputs. QiT uses the explicit cosine feature kernel at the bottom; it does not generally approximate the circuit-dependent fidelity kernel at the top.}
\label{fig:hilbert_commutative}
\end{figure}
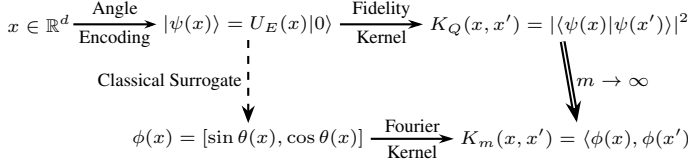

Before projection, $\phi(h_i)^\top\phi(h_j)$ is the cosine kernel above. After applying independently learned query and key projections, $q_i^\top k_j$ is a learned bilinear similarity in that feature space; it need not be symmetric or positive semidefinite and should not be interpreted as quantum-state fidelity. Figure~\ref{fig:hilbert_commutative} contrasts these two kernel constructions. For all heads, feature projection costs $\mathcal{O}(BLQd)$ and attention costs $\mathcal{O}(BL^2d)$.

\paragraph{Scope of the circuit interpretation.}

A circuit-based attention mechanism could encode a token as a normalized state
\begin{equation}
|\psi(h)\rangle=U_E(h)|0\rangle,
\qquad \langle\psi(h)|\psi(h)\rangle=1,
\end{equation}


and to form query and key states $|\psi_Q(h_i)\rangle=U_Q|\psi(h_i)\rangle$ and $|\psi_K(h_j)\rangle=U_K|\psi(h_j)\rangle$, where  $U_Q(\theta_Q)$ and $U_K(\theta_K)$ be unitary families. Their fidelity score and normalized attention weight would be
\begin{equation}
s_{ij} = \big|\langle \psi_Q(h_i) \mid \psi_K(h_j) \rangle \big|^2 \in [0,1].
\end{equation}
Then for any temperature $\tau>0$, the attention weights
\begin{equation}
A_{ij} = \frac{\exp(s_{ij}/\tau)}{\sum_{j'} \exp(s_{ij'}/\tau)}
\end{equation}
form a valid probability distribution over $j$ and induce an attention operator
\begin{equation}
\mathrm{Attn}(h_i) = \sum_{j=1}^{L} A_{ij} v_j.
\end{equation}

 Moreover, $s_{ij}$ defines a positive-definite kernel over token embeddings.
Unitarity of $U_E$, $U_Q$, and $U_K$ ensures all states are normalized. Unitarity guarantees normalized states,Hence $s_{ij} \in [0,1]$ and $\exp(s_{ij}/\tau)>0$,  and softmax guarantees $A_{ij}>0$ and  so $\sum_j A_{ij}=1$. When the same state map is used on both arguments, fidelity is a positive-semidefinite kernel because $|\langle\psi_i|\psi_j\rangle|^2=\operatorname{tr}(\rho_i\rho_j)$ for $\rho_i=|\psi_i\rangle\langle\psi_i|$. With distinct query and key maps, however, $s_{ij}$ need not be symmetric and therefore is not generally a kernel.

QiT neither estimates these fidelities nor approximates them through an overlap routine such as a swap test. It uses the finite real feature map $\phi$ and learns task-dependent query and key projections before softmax attention. The shared periodic dependence on encoded angles motivates the design, while the learned projections allow the visual recognition objective to determine the similarity. QiT attention costs $\mathcal{O}(BL^2d_h)$ per head, as in scaled dot-product attention. Its empirical behavior is therefore evidence about periodic classical features, not quantum-state fidelity or quantum advantage.

\begin{figure}[t]
\centering
\footnotesize
\begin{tikzpicture}[
    >=Stealth,
    thick,
    node distance=1.1cm,
    every node/.style={draw, rectangle, rounded corners, inner sep=3pt}
]

\node (rx) {RX($\theta_1$)};
\node (rz) [right=of rx] {RZ($\theta_2$)};
\node (cnot) [right=of rz] {CNOT};
\node (meas) [right=of cnot] {Measurement};

\node (lin1) [below=1.1cm of rx] {Linear $W_1$};
\node (act)  [right=of lin1] {Nonlinearity};
\node (gate) [right=of act] {Gated Mix};
\node (lin2) [right=of gate] {Linear $W_2$};

\draw[->] (rx) -- (rz);
\draw[->] (rz) -- (cnot);
\draw[->] (cnot) -- (meas);

\draw[->] (lin1) -- (act);
\draw[->] (act) -- (gate);
\draw[->] (gate) -- (lin2);

\draw[<->, dashed]
  (rx.south) -- (lin1.north)
  node[midway, right] {\scriptsize Angle encoding};

\draw[<->, dashed]
  (cnot.south) -- (gate.north)
  node[midway, left] {\scriptsize Entangle $\leftrightarrow$ Gating};

\draw[<->, dashed]
  (meas.south) -- (lin2.north)
  node[midway, left] {\scriptsize Readout};

\end{tikzpicture}
\vspace{-0.5em}
\caption{Design analogy between a parameterized circuit and the QiT channel mixer. The correspondence motivates the sequence of operations but does not imply equivalent states or functions.}
\label{fig:pqc_to_transformer}
\end{figure}
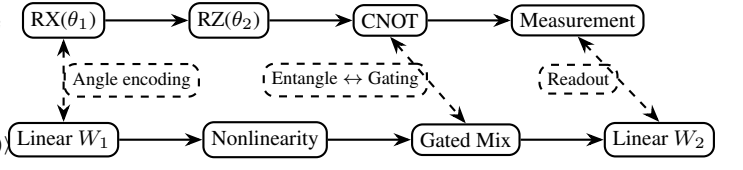


\begin{table*}[t]
\centering
\scriptsize
\caption{One-epoch CIFAR simulator diagnostic with patch size 4, one block, and batch size 2. $Q{=}4$ denotes learned angles for QiT and simulated qubits for FQT-P. FLOPs are recorded for CViT and QiT; FQT-P is listed symbolically because its cost depends on circuit and simulator details. Accuracy values are single-run, non-converged measurements.FQT-P~=~Full Quantum Transformer in PennyLane.}
\label{tab:main_results}
\vspace{-0.9em}
\begin{tabular}{lcccccccc}
\toprule
\textbf{Model} & \textbf{Classes} & \textbf{Angles/Qubits} & \textbf{Params} & \textbf{FLOPs} & \textbf{Energy (J)} & \textbf{Acc. (\%)} & \textbf{Latency (s)} & \textbf{Wall-clock (s)} \\
\midrule
CViT  & 2  & 0 & 147{,}458 & 8{,}996{,}352 & 9.60 & 50.00 & 0.0010 & 63.72 \\
QiT   & 2  & 4 angles & 324{,}690 & 20{,}756{,}672 & 10.76 & 86.35 & 0.0016 & 102.91 \\
FQT-P & 2  & 4 qubits & 16{,}394 & $\mathcal{O}(N^2 2^Q)$ & 23{,}904.70 & 81.20 & 2.4987 & 90{,}745.75 \\
\midrule
CViT  & 4  & 0 & 147{,}716 & 8{,}996{,}608 & 9.40 & 25.00 & 0.0008 & 127.97 \\
QiT   & 4  & 4 angles & 324{,}948 & 20{,}756{,}928 & 10.64 & 71.98 & 0.0015 & 202.02 \\
FQT-P & 4  & 4 qubits & 16{,}652 & $\mathcal{O}(N^2 2^Q)$ & 13{,}697.42 & 62.42 & 2.3363 & 178{,}606.09 \\
\midrule
CViT  & 10 & 0 & 147{,}490 & 8{,}997{,}376 & 5.60 & 10.00 & 0.0007 & 253.50 \\
QiT   & 10 & 4 angles & 325{,}722 & 20{,}757{,}696 & 6.89 & 50.39 & 0.0013 & 447.23 \\
FQT-P & 10 & 4 qubits & 17{,}426 & $\mathcal{O}(N^2 2^Q)$ & 13{,}641.24 & 32.33 & 2.3174 & 445{,}765.80 \\
\bottomrule
\end{tabular}
\end{table*}


\begin{table*}[t]
\centering
\caption{Unified asymptotic comparison of Transformer variants. $N$~=~tokens, $D$~=~embedding dim, $Q$~=~qubits, $L$~=~circuit depth. Runtime slowdown is relative to CViT.}
\label{tab:unified_comparison}
\vspace{-0.9em}
\resizebox{\textwidth}{!}{
\begin{tabular}{lcccccc}
\toprule
\textbf{Model} & \textbf{Token mixing (attn.)} & \textbf{Channel mixer} & \textbf{Total complexity} & \textbf{Slowdown} & \textbf{Feasibility (CIFAR)} & \textbf{Acc} \\
\midrule
CViT                  & $\mathcal{O}(N^2 D)$            & $\mathcal{O}(N D^2)$    & $\mathcal{O}(N^2 D)$            & $1\times$         & Fully trainable           & 0.50 \\
QiT (surrogate)       & $\mathcal{O}(N^2 D)$            & $\mathbf{\mathcal{O}(Q L)}$ & $\mathcal{O}(N^2 D)$            & $1.1\times$       & Fully trainable           & 0.86 \\
FQT-P (PennyLane)     & $\mathbf{\mathcal{O}(N^2 \cdot 2^Q)}$ & $\mathcal{O}(Q L)$ & $\mathbf{\mathcal{O}(N^2 \cdot 2^Q)}$ & $\sim 100\times$  & Slow but feasible         & 0.81 \\
FQT-Q (Qiskit)        & $\mathbf{\mathcal{O}(N^2 \cdot 2^Q)}$ & $\mathcal{O}(Q L)$ & $\mathbf{\mathcal{O}(N^2 \cdot 2^Q)}$ & $\sim 1000\times$ & Infeasible at CIFAR scale & -- \\
\bottomrule
\end{tabular}}
\end{table*}

\subsection{Quantum-Inspired Gated Channel Mixer}

\subsubsection{Design rationale}

Parameterized quantum circuits use two types of operations: single-qubit rotation gates that depend on learnable parameters, and multi-qubit entangling gates like CNOT that create non-separable correlations between qubits. A quantum circuit with parameters $\Theta = \{\theta_1,\ldots,\theta_p\}$ computes:
\begin{equation}
f_{\Theta}(x)
=
\langle 0 | U_{\Theta}^\dagger(x)\, O \, U_{\Theta}(x) | 0 \rangle,
\end{equation}
where $x$ is the classical input, $U_{\Theta}(x)$ applies encoding and variational gates, and $O$ is a measurement observable. Entangling gates can produce states and measurement correlations that cannot be factored into independent single-qubit contributions. Motivated by this sequence of rotation, interaction, and readout, we use a classical gated block. It introduces multiplicative feature interactions without representing a quantum state.

\subsubsection{Gated multiplicative mixing}

Given a token representation $X$, the channel mixer applies
\begin{align}
H_1 &= \mathrm{GELU}(X W_1), \\
G   &= \sigma(X W_g), \\
\mathrm{Mix}(X) &= (H_1 \odot G) W_2,
\end{align}
where $\odot$ denotes elementwise multiplication, $W_1$ and $W_g$ map to hidden width $r$, and $W_2$ projects back to width $d$. The Hadamard product creates input-dependent, non-additive feature interactions. Such products are qualitatively reminiscent of interaction terms in circuit expectation values, but no one-to-one correspondence to an entangling gate is assumed. The computational cost is $\mathcal{O}(BLdr)$, matching a gated Transformer feed-forward block of hidden width $r$.

\subsubsection{Architectural interpretation}

Figure~\ref{fig:pqc_to_transformer} illustrates the mapping between quantum circuit elements and QiT components. Angle encoding emulates single-qubit rotations ($R_X$, $R_Z$), gated mixing emulates two-qubit gates (CNOT), and multi-head attention corresponds to parallel measurements. Residual connections act as classical analogs of quantum error mitigation by providing gradient highways that stabilize training.  Angle encoding motivates periodic features, multiplicative gating supplies feature interactions, and a linear layer provides readout. Residual connections and normalization follow standard Transformer practice. Because QiT uses a finite classical feature map, it does not inherit the exponential state dimension or the full function class of a quantum circuit.




\section{Experiments}

We organize the evaluation around three questions. \textbf{(Q1) Simulator cost:} under one controlled CIFAR protocol, how does a four-qubit PennyLane implementation compare with classical execution? \textbf{(Q2) Portability:} can one QiT configuration train across datasets with different resolutions and class counts? \textbf{(Q3) ImageNet scale:} can larger QiT variants be trained on ImageNet-1K, and where do their accuracy and computational cost lie relative to published vision architectures? These experiments evaluate a classical model and do not test quantum advantage.

\subsection{Experimental Setup}

\paragraph{Datasets.}
The portability study comprises eight datasets. CIFAR-10, SVHN, and CIFAR-100 use $32\times32$ images; STL-10 uses $96\times96$ images; Oxford-IIIT Pets, Food-101, Flowers-102, and Caltech-256 use $224\times224$ images. They span 10 to 257 classes and 3,680 to 75,750 training images, as detailed in Table~\ref{tab:dataset_summary}. ImageNet-1K is evaluated separately with 1.2M training images, 1,000 classes, and $224\times224$ inputs. The simulator diagnostic uses reduced 2-, 4-, and 10-class CIFAR tasks; these are distinct from the 300-epoch CIFAR-10 result in the portability study.

\paragraph{Protocols.}
The three studies use separate settings. For Q1, CViT, QiT, and FQT-P use patch size 4, one block, batch size 2, and one training epoch; QiT and FQT-P use $Q=4$, where $Q$ denotes learned angles for QiT and qubits for FQT-P. For Q2, a fixed QiT architecture uses patch size 4, depth 12, width 256, four attention heads, and $Q=16$ learned angles. It is trained for 300 epochs with AdamW, cosine learning-rate decay, random cropping, horizontal flipping, and color jitter on four Tesla T4 GPUs; per-GPU batch sizes appear in Table~\ref{tab:dataset_summary}. For Q3, the QiT-XS/S/B variants use the parameter and FLOP budgets reported in Table~\ref{tab:similar_arch_imagenet}. Results must be interpreted within each protocol rather than compared across protocols.

\paragraph{Reporting scope.}
All tables report point estimates from the available runs. The current records do not include repeated seeds, uncertainty intervals, complete optimizer hyperparameters, reduced-class CIFAR split definitions, or the energy-measurement procedure. We therefore use the results to document observed accuracy, runtime, and scalability, not statistical superiority. A final reproducible submission should provide these missing details and repeated-run confidence intervals.

\begin{table}[h!]
\centering
\caption{ImageNet-1K top-1 accuracy, GFLOPs,  and Parameters are reported for representative architectures. Results are drawn from their respective papers and use different training recipes; the table provides scale and accuracy context rather than controlled pairwise comparisons. QiT rows are highlighted.}
\label{tab:similar_arch_imagenet}
\vspace{-0.9em}
\begin{tabular}{l|c|c|c|c}
\toprule
Network & Size & Params & FLOPs & Top-1 \\
\midrule
\multicolumn{5}{c}{CNN}\\
\midrule
ResNet-50~\citep{he2016deep}            & $224^2$ & 25.5M & 4.1G  & 78.3 \\
ResNeXt101~\citep{xie2017aggregated}    & $224^2$ & 83.5M & 15.6G & 81.5 \\
EfficientNet-B5~\citep{tan2019efficientnet} & $456^2$ & 30M & 9.9G & 83.6 \\
\midrule
\multicolumn{5}{c}{Transformer-MLP Mixer}\\
\midrule
gMLP-Ti~\cite{NEURIPS2021_4cc05b35}     & $224^2$ & 6M  & 1.4G  & 72.0 \\
gMLP-S~\cite{NEURIPS2021_4cc05b35}      & $224^2$ & 20M & 4.5G  & 79.4 \\
gMLP-B~\cite{NEURIPS2021_4cc05b35}      & $224^2$ & 73M & 15.8G & 81.6 \\
ResMLP-12~\cite{touvron2022resmlp}      & $224^2$ & 15M & 3.0G  & 76.6 \\
ResMLP-36~\cite{touvron2022resmlp}      & $224^2$ & 45M & 8.9G  & 79.7 \\
\midrule
\multicolumn{5}{c}{Transformer-Spectral}\\
\midrule
Fnet-S~\cite{lee2021fnet}               & $224^2$ & 15M & 2.9G  & 71.2 \\
GFNet-Ti~\cite{rao2021global}           & $224^2$ & 7M  & 1.3G  & 74.6 \\
GFNet-XS~\cite{rao2021global}           & $224^2$ & 16M & 2.9G  & 78.6 \\
GFNet-S~\cite{rao2021global}            & $224^2$ & 25M & 4.5G  & 80.0 \\
SpectFormer-XS~\cite{patro2025spectformer} & $224^2$ & 20M & 4.0G & 80.2 \\
SpectFormer-S~\cite{patro2025spectformer}  & $224^2$ & 32M & 6.6G & 81.7 \\
SpectFormer-B~\cite{patro2025spectformer}  & $224^2$ & 57M & 11.5G & 82.1 \\
\midrule
\multicolumn{5}{c}{Transformer-State-Space}\\
\midrule
Vim-Ti~\citep{zhu2024vision}            & $224^2$ & 7M    & 1.5G  & 76.0 \\
Vim-S~\citep{zhu2024vision}             & $224^2$ & 26M   & 5.1G  & 80.4 \\
LocalVMamba-T~\citep{huang2024localmamba} & $224^2$ & 26.0M & 5.7G  & 82.7 \\
LocalVMamba-S~\citep{huang2024localmamba} & $224^2$ & 50.0M & 11.4G & 83.7 \\
SiMBA-S~\citep{patro2024simba}          & $224^2$ & 15.3M & 2.4G  & 81.7 \\
SiMBA-B~\citep{patro2024simba}          & $224^2$ & 22.8M & 4.2G  & 83.5 \\
VMamba-T~\citep{liu2024vmamba}          & $224^2$ & 31.0M & 4.9G  & 82.5 \\
VMamba-S~\citep{liu2024vmamba}          & $224^2$ & 50.0M & 8.7G  & 83.6 \\
\midrule
\multicolumn{5}{c}{Transformer-Attention}\\
\midrule
ViT-B/16~\cite{dosovitskiy2020image}    & $384^2$ & 86M & 55.5G & 77.9 \\
DeiT-Ti~\cite{touvron2021training}      & $224^2$ & 5M  & 1.2G  & 72.2 \\
DeiT-S~\cite{touvron2021training}       & $224^2$ & 22M & 4.6G  & 79.8 \\
DeiT-B~\cite{touvron2021training}       & $224^2$ & 86M & 17.5G & 81.8 \\
\midrule
\multicolumn{5}{c}{Transformer-Quantum-Inspired (ours)}\\
\midrule
\rowcolor{gray!15}QiT-XS & $224^2$ & 18.1M & 3.8G  & 75.0 \\
\rowcolor{gray!15}QiT-S  & $224^2$ & 31.0M & 5.6G  & 77.4 \\
\rowcolor{gray!15}QiT-B  & $224^2$ & 45.7M & 11.5G & 78.3 \\
\bottomrule
\end{tabular}
\vspace{-1.5em}
\end{table} 

\begin{table*}[t]
\centering
\caption{Single-run QiT results across eight image datasets. All runs use four Tesla T4 GPUs, patch size 4, depth 12, width 256, four heads, $Q{=}16$ learned angles, and 300 epochs. Batch is per GPU; latency is reported per image.}
\label{tab:dataset_summary}
\vspace{-0.9em}
\resizebox{\textwidth}{!}{
\begin{tabular}{@{}lcccccccccc@{}}
\toprule
\textbf{Dataset} & \textbf{Classes} & \textbf{Image Size} & \textbf{Train} & \textbf{Batch/GPU} & \textbf{Epochs} & \textbf{Params (M)} & \textbf{GFLOPs} & \textbf{Acc (\%)} & \textbf{Wall clock} & \textbf{Inference (ms)} \\
\midrule
CIFAR-10    & 10  & $32{\times}32$   & 50{,}000  & 128 & 300 & 5.48 & 0.355 & 78.84 & $\sim$1.74h  & 14.50 \\
SVHN        & 10  & $32{\times}32$   & 73{,}257  & 128 & 300 & 5.48 & 0.355 & 95.94 & $\sim$2.55h  & 14.66 \\
STL-10      & 10  & $96{\times}96$   & 5{,}000   & 32  & 300 & 5.61 & 3.386 & 63.50 & $\sim$2.07h  & 14.65 \\
Pets        & 37  & $224{\times}224$ & 3{,}680   & 4   & 300 & 6.27 & 24.58 & 71.23 & $\sim$20.93h & 77.18 \\
CIFAR-100   & 100 & $32{\times}32$   & 50{,}000  & 128 & 300 & 5.50 & 0.355 & 51.52 & $\sim$1.75h  & 14.39 \\
Food-101    & 101 & $224{\times}224$ & 75{,}750  & 4   & 300 & 6.29 & 24.58 & 72.00 & $\sim$78.34h & 80.12 \\
Flowers-102 & 102 & $224{\times}224$ & 6{,}149   & 4   & 300 & 6.29 & 24.65 & 65.00 & $\sim$6.77h  & 79.31 \\
Caltech-256 & 257 & $224{\times}224$ & 24{,}000  & 4   & 300 & 6.33 & 24.58 & 68.23 & $\sim$23.74h & 79.98 \\
\bottomrule
\end{tabular}}
\end{table*}

\subsection{Comparison with Simulated Quantum Transformers}

Table~\ref{tab:main_results} compares a classical Vision Transformer (CViT), QiT, and a four-qubit Full Quantum Transformer simulated in PennyLane (FQT-P) under the Q1 protocol. In this table, ``angles/qubits'' distinguishes QiT's four learned angles from FQT-P's four simulated qubits.

\paragraph{Runtime and latency.}
Across 2, 4, and 10 classes, FQT-P requires 90,745.75, 178,606.09, and 445,765.80 seconds, respectively, versus 102.91, 202.02, and 447.23 seconds for QiT. The corresponding FQT-P/QiT wall-clock ratios are $881.8\times$, $884.1\times$, and $996.7\times$. Reported inference latency shows a similarly large gap of $1557.5$--$1782.6\times$. QiT is not cost-free relative to CViT: it uses about $2.30\times$ as many recorded FLOPs and $2.20\times$ as many parameters in this setup, yielding a $1.58$--$1.76\times$ wall-clock increase. Thus the pilot demonstrates a large implementation-cost gap between classical tensor execution and this PennyLane simulation, while also quantifying QiT's overhead over the small CViT.

\paragraph{Accuracy and interpretation.}
After one epoch, QiT records 86.35\%, 71.98\%, and 50.39\% accuracy for 2, 4, and 10 classes; FQT-P records 81.20\%, 62.42\%, and 32.33\%. CViT remains exactly at chance (50\%, 25\%, and 10\%), indicating that this short protocol is unsuitable for assessing its converged performance. The observed QiT--FQT-P differences are therefore optimization snapshots, not evidence of architectural superiority. Parameter counts are also unmatched: QiT has roughly 325K parameters, CViT 147K, and FQT-P 16--17K. The defensible conclusion is computational: the tested simulator does not provide a practical route to the longer and larger studies below.

\paragraph{Energy measurements.}
FQT-P's recorded energy is three orders of magnitude above the classical runs, but the measurement procedure, simulator implementation, and device utilization are undocumented. We report these values for completeness and draw no hardware-independent energy-efficiency conclusion from them.

\paragraph{Asymptotic complexity analysis.}
Table~\ref{tab:unified_comparison} separates asymptotic complexity from measured runtime. QiT's periodic feature projection costs $\mathcal{O}(NQD)$ per layer, attention costs $\mathcal{O}(N^2D)$, and a gated mixer of hidden width $r$ costs $\mathcal{O}(NDr)$. Its total layer cost is therefore $\mathcal{O}(N^2D+NQD+NDr)$, retaining quadratic dependence on token count. CViT additionally incurs its channel-MLP cost $\mathcal{O}(ND^2)$. FQT-P complexity is circuit- and simulator-dependent; a dense state-vector simulator stores $2^Q$ amplitudes, but the measured slowdown cannot be generalized to other simulation methods or quantum hardware. We omit the proposed Qiskit row because no Qiskit experiment is recorded in the available results.

\subsection{Context from Quantum Attention Models}
Table~\ref{tab:quantum_complexity} summarizes reported resource characteristics for QSANN~\cite{li2023quantumselfattentionneuralnetworks}, QSAM~\cite{qsam2023}, HQViT~\cite{hqvit2025}, QViT~\cite{cherrat2024qvt}, QKSAN~\cite{Zhao_2024_QKSAN}, QSAN~\cite{shi2024qsan}, and F-QSANN~\cite{zheng2023fqsann}. Circuit architecture, encoding, measurements, and hardware assumptions differ across these works, so the entries provide qualitative context rather than a normalized runtime ranking. QiT has zero quantum-resource requirements by construction.

\subsection{Scalability to ImageNet}

Table~\ref{tab:similar_arch_imagenet} places QiT-XS/S/B alongside published ImageNet-1K results. Within the QiT family, increasing from XS to S raises top-1 accuracy from 75.0\% to 77.4\% while increasing compute from 3.8 to 5.6 GFLOPs. Scaling from S to B adds 14.7M parameters and 5.9 GFLOPs for a further 0.9 percentage-point gain, reaching 78.3\%. This monotonic but diminishing gain indicates that the architecture remains trainable as capacity grows, although it does not establish optimal scaling.

Across published models, QiT-B matches the listed ResNet-50 accuracy and exceeds the listed ViT-B/16 result by 0.4 points with fewer reported FLOPs, but these are not controlled comparisons because input resolution, augmentation, optimization, and pretraining may differ. QiT trails DeiT-B by 3.5 points and the strongest spectral and state-space entries by 3.8--5.4 points. Accordingly, Table~\ref{tab:similar_arch_imagenet} supports ImageNet-scale feasibility and locates the current accuracy--cost trade-off; it does not support a state-of-the-art claim.

\subsection{Results Across Multiple Benchmarks}

Table~\ref{tab:dataset_summary} evaluates one QiT configuration without dataset-specific changes to depth, width, head count, or angle count. For $32\times32$ inputs, the model uses 5.48--5.50M parameters and 0.355 GFLOPs, obtaining 78.84\% on CIFAR-10, 95.94\% on SVHN, and 51.52\% on CIFAR-100. The spread across these values reflects different tasks and class structures, so it should not be interpreted as a direct ranking of dataset difficulty. Reported latency is consistent within this resolution group at 14.39--14.66 ms per image.

Increasing resolution changes the resource profile more than the classifier size. STL-10 at $96\times96$ requires 3.386 GFLOPs and reaches 63.50\%. At $224\times224$, the four datasets require 24.58--24.65 GFLOPs and 6.27--6.33M parameters, with accuracy ranging from 65.00\% on Flowers-102 to 72.00\% on Food-101. Their reported latency lies between 77.18 and 80.12 ms per image. Relative to the $32\times32$ runs, the $224\times224$ configuration uses about $69\times$ more FLOPs but approximately $5.4\times$ the measured latency, consistent with different levels of GPU utilization across input sizes rather than proportional latency scaling.

Wall-clock time varies with both image resolution and training-set size: Food-101 is the longest run at approximately 78.34 hours, while the smaller $32\times32$ datasets finish in under 2.6 hours. Taken together, the table establishes that the same QiT block can be optimized across substantial changes in resolution, sample count, and class count. Because no matched baseline or component ablation is reported under this 300-epoch protocol, the table cannot attribute accuracy gains to periodic encoding or gated mixing.

\paragraph{Experimental findings.}
The experiments answer the three evaluation questions as follows. For Q1, the tested four-qubit PennyLane implementation is $882$--$997\times$ slower in wall-clock time than QiT, while QiT is $1.58$--$1.76\times$ slower than the small CViT. For Q2, one QiT configuration trains across eight datasets spanning $32\times32$ to $224\times224$ inputs and 10 to 257 classes. For Q3, QiT scales to ImageNet-1K and reaches 78.3\% top-1 accuracy, but remains below stronger contemporary backbones in the contextual table. The current evidence establishes execution feasibility and measured cost; it does not establish that the quantum-motivated components outperform a converged, parameter-matched Transformer. That question requires controlled ablations, repeated seeds, and a complete shared training protocol.

\section{Conclusion}

We presented QiT, a classical vision Transformer that forms attention projections from learned periodic token features and uses gated multiplicative channel mixing. The construction is motivated by angle-encoded quantum models but does not simulate a general quantum circuit or claim quantum advantage. QiT scales across eight image datasets and reaches 78.3\% ImageNet-1K top-1 accuracy with 45.7M parameters. A controlled four-qubit pilot further shows the high cost of the tested PennyLane simulator relative to ordinary tensor operations. The current evidence establishes scalability and motivates a focused study of periodic feature maps in attention; controlled component ablations and repeated runs remain necessary to determine their advantage over matched classical alternatives.

{
    \small
    \bibliographystyle{ieeenat_fullname}
    \bibliography{main}
}

\appendix
\section{Supplementary Material}
\label{sec:supplementary}

This supplement provides implementation details and derivations for the model in the main paper. QiT is implemented entirely with classical tensor operations. The terms ``angle,'' ``Hilbert space,'' and ``circuit-inspired'' describe the motivation for its parameterization; they do not imply execution on quantum hardware, simulation of a general quantum circuit, or quantum computational advantage.

\subsection{QiT Encoder Block}
\label{sec:supp_algorithm}

Algorithm~\ref{alg:qit_encoder_supp} gives the operations in one encoder block. The input $H$ contains patch and positional embeddings. The periodic feature map is shared by the query, key, and value projections, while each attention head has separate learned projection matrices. Standard dropout operations are omitted from the notation.

\begin{algorithm}[t]
\caption{One QiT encoder block}
\label{alg:qit_encoder_supp}
\begin{algorithmic}[1]
\Require Token representations $H\in\mathbb{R}^{B\times N\times d}$; number of learned angles $Q$
\Ensure Updated representations $H_{\mathrm{out}}\in\mathbb{R}^{B\times N\times d}$
\State $\Theta \leftarrow \pi\tanh(HW_{\theta}+b_{\theta})$
\State $\Phi \leftarrow Q^{-1/2}[\sin(\Theta),\cos(\Theta)]$
\For{each attention head $a$}
    \State $Q_a\leftarrow\Phi W_Q^{(a)}$, $K_a\leftarrow\Phi W_K^{(a)}$, $V_a\leftarrow\Phi W_V^{(a)}$
    \State $A_a\leftarrow\operatorname{softmax}(Q_aK_a^{\top}/\sqrt{d_h})$
    \State $Z_a\leftarrow A_aV_a$
\EndFor
\State $Z\leftarrow\operatorname{Concat}(Z_1,\ldots,Z_{n_h})W_O$
\State $H'\leftarrow\operatorname{LayerNorm}(H+Z)$
\State $U\leftarrow\operatorname{GELU}(H'W_1)$
\State $G\leftarrow\sigma(H'W_g)$
\State $M\leftarrow(U\odot G)W_2$
\State $H_{\mathrm{out}}\leftarrow\operatorname{LayerNorm}(H'+M)$
\State \Return $H_{\mathrm{out}}$
\end{algorithmic}
\end{algorithm}

The block differs from a standard Transformer encoder in two places. First, query, key, and value projections operate on $\Phi$ rather than directly on $H$. Second, the usual channel MLP is replaced by the gated multiplicative map $(U\odot G)W_2$. Residual connections, normalization, multi-head composition, and softmax attention remain standard Transformer operations.

\subsection{Exact Property of the Periodic Feature Map}
\label{sec:supp_kernel}

For a token representation $h$, let
\begin{equation}
\theta(h)=\pi\tanh(hW_{\theta}+b_{\theta}),\qquad
\phi(h)=\frac{1}{\sqrt Q}[\sin\theta(h),\cos\theta(h)].
\end{equation}
The feature vector has unit norm because
\begin{equation}
\|\phi(h)\|_2^2
=\frac{1}{Q}\sum_{q=1}^{Q}\left(\sin^2\theta_q(h)+\cos^2\theta_q(h)\right)=1.
\end{equation}
Its inner product is the finite cosine kernel
\begin{align}
k(h,h')
&=\phi(h)^\top\phi(h')\\
&=\frac{1}{Q}\sum_{q=1}^{Q}
\cos\!\left(\theta_q(h)-\theta_q(h')\right).
\end{align}
For arbitrary inputs $\{h_i\}_{i=1}^{m}$ and coefficients $\{c_i\}_{i=1}^{m}$,
\begin{equation}
\sum_{i,j=1}^{m}c_ic_jk(h_i,h_j)
=\left\|\sum_{i=1}^{m}c_i\phi(h_i)\right\|_2^2\geq0.
\end{equation}
Thus, $k$ is positive semidefinite. This statement applies to the unprojected feature map. QiT uses
\begin{equation}
q_i^\top k_j
=\phi(h_i)^\top W_QW_K^\top\phi(h_j)
\end{equation}
inside attention. Since $W_Q$ and $W_K$ are learned independently, this score is generally asymmetric and is not, in general, a positive-semidefinite kernel.

For comparison, independent single-qubit $R_Y$ encodings give the product-state fidelity
\begin{equation}
K_{\mathrm{prod}}(h,h')
=\prod_{q=1}^{Q}\cos^2\!\left(
\frac{\theta_q(h)-\theta_q(h')}{2}\right).
\end{equation}
This product is different from QiT's average cosine kernel. Entangling circuits introduce additional circuit-dependent terms. The shared periodic dependence on encoded angles motivates QiT's feature map, but it does not establish equivalence to a quantum fidelity kernel.

\subsection{Computational Cost}
\label{sec:supp_complexity}

For batch size $B$, $N$ tokens, embedding width $d$, $Q$ learned angles, and gated-mixer width $r$, angle projection and feature construction require $\mathcal{O}(BNdQ)$ operations and store $2Q$ features per token. Projection to the attention heads requires $\mathcal{O}(BNQd)$ operations. Scaled dot-product attention requires $\mathcal{O}(BN^2d)$ operations, and the gated channel mixer requires $\mathcal{O}(BNdr)$. The resulting per-block cost is
\begin{equation}
\mathcal{O}\!\left(BN^2d+BNQd+BNdr\right).
\end{equation}
QiT therefore retains the quadratic dependence on token count of standard softmax attention. These are classical operation counts; they are not quantum runtime bounds.

\subsection{Experimental Protocol and Interpretation}
\label{sec:supp_protocol}

The reported results come from three separate protocols and should be interpreted within their respective settings.

\paragraph{Simulator diagnostic.}
CViT, QiT, and the PennyLane model use patch size 4, one encoder block, batch size 2, and one training epoch on reduced 2-, 4-, and 10-class CIFAR tasks. QiT uses four learned angles, whereas the PennyLane model uses four simulated qubits. The models are not parameter matched, and one epoch does not measure converged accuracy. This experiment supports only the reported implementation-cost comparison for the tested configurations.

\paragraph{Multi-dataset study.}
The eight-dataset study uses one QiT configuration with patch size 4, depth 12, width 256, four attention heads, and 16 learned angles. Each model is trained for 300 epochs with AdamW, cosine learning-rate decay, random cropping, horizontal flipping, and color jitter on four Tesla T4 GPUs. Batch size varies with image resolution as reported in the main paper.

\paragraph{ImageNet-1K study.}
QiT-XS, QiT-S, and QiT-B are evaluated at $224\times224$ resolution with the parameter counts and FLOPs reported in the main paper. Comparisons with published architectures provide accuracy--cost context rather than controlled pairwise evidence because their training recipes can differ.

All experimental values are point estimates from the available runs. Repeated seeds, uncertainty intervals, complete optimizer settings, reduced-class split definitions, and the energy-measurement procedure are not available in the current records. Accordingly, the evidence establishes that QiT can be trained at image-classification scale and documents the measured cost of the tested implementations. It does not establish statistical superiority over a converged, parameter-matched Transformer or any quantum advantage.

\end{document}